\documentclass[sigconf]{acmart}

\usepackage{amsmath,amssymb,amsfonts}
\usepackage{algorithmic}
\usepackage{textcomp}
\usepackage{multirow}
\usepackage{makecell}
\usepackage{csquotes}
\usepackage{float}
\usepackage[breakable]{tcolorbox}
\usepackage{enumitem}
\usepackage{array}
\usepackage{url}
\usepackage{dblfloatfix}
\usepackage[capitalise]{cleveref}

\newcolumntype{L}[1]{>{\raggedright\let\newline\\\arraybackslash\hspace{0pt}}m{#1}}
\newcolumntype{C}[1]{>{\centering\let\newline\\\arraybackslash\hspace{0pt}}m{#1}}
\newcolumntype{R}[1]{>{\raggedleft\let\newline\\\arraybackslash\hspace{0pt}}m{#1}}

\renewcommand{\arraystretch}{1.5}
\newcommand{\code}[1]{\texttt{#1}}
\newcommand{\myparagraph}[1]{\smallskip\noindent\textbf{#1}\hspace{0.5em}}
\newcommand{\takeaway}[1]{
\begin{tcolorbox}[
  leftrule=0.5mm, rightrule=0.5mm, toprule=0.5mm, bottomrule=0.5mm,
  left=2pt, right=2pt, top=2pt, bottom=2pt,
  breakable
  ]%
\em #1
\end{tcolorbox}
}

\begin{document}

\title{Text Tells the Cost: Predicting and Analyzing Repayment Effort of Self-Admitted Technical Debt}

\author{Yikun Li}
\affiliation{%
  \institution{Bernoulli Institute for Mathematics, Computer Science and Artificial Intelligence, University of Groningen}
  \streetaddress{Broerstraat 5}
  \postcode{9712 CP}
  \city{Groningen}
  \country{The Netherlands}
}

\author{Mohamed Soliman}
\affiliation{%
  \institution{Bernoulli Institute for Mathematics, Computer Science and Artificial Intelligence, University of Groningen}
  \streetaddress{Broerstraat 5}
  \postcode{9712 CP}
  \city{Groningen}
  \country{The Netherlands}
}

\author{Paris Avgeriou}
\affiliation{%
  \institution{Bernoulli Institute for Mathematics, Computer Science and Artificial Intelligence, University of Groningen}
  \streetaddress{Broerstraat 5}
  \postcode{9712 CP}
  \city{Groningen}
  \country{The Netherlands}
}

\author{Jie Tan}
\email{j.tanjie@outlook.com}
\affiliation{%
  \institution{Intelligent Game and Decision Lab}
  \city{Beijing}
  \country{China}
}

\author{Jiakun Liu}
\email{jiakunliu@hit.edu.cn}
\affiliation{%
  \institution{Harbin Institute of Technology}
  \city{Harbin}
  \country{China}
}

\setcopyright{none}
\settopmatter{printacmref=false}
\renewcommand\footnotetextcopyrightpermission[1]{}

\begin{abstract}
Self-Admitted Technical Debt (SATD) is technical debt explicitly documented by developers in software artifacts such as source code comments and commit messages.
As SATD can hinder software development and maintenance, estimating the effort required to repay it is crucial for effective prioritization. However, we currently lack both an understanding of SATD repayment effort and automated approaches to estimate it from textual descriptions.

To bridge this gap, we curate a comprehensive dataset of 341,740 SATD items from 2,568,728 commits across 1,060 Apache repositories and propose \underline{P}redicting \underline{R}epayment \underline{E}ffort of \underline{S}ATD using \underline{T}extual \underline{I}nformation (PRESTI).
We analyze the repayment effort comparing SATD vs.\ non-SATD items and across different SATD types, evaluate machine learning models for automated effort prediction, and identify keywords associated with varying effort levels.

Our findings show that SATD and non-SATD items require similar direct resolution effort (mean total lines changed: 131.5 vs.\ 132.7), but SATD creates significantly larger ripple effects (effect size = 0.08, p $<$ 0.05). Different SATD types demand distinct effort levels: documentation debt requires the least effort (mean 72.7 lines), even lower than non-SATD items (132.7), while requirement debt demands the most (169.7).
BERT- and TextCNN-based models outperform traditional machine learning methods and the baseline in estimating repayment effort, with BERT achieving the best RMSE of 13.5 and reducing the baseline error (38.1) by 64.6\%.
We further identify interpretable keywords associated with low- and high-effort repayments, providing actionable insights for SATD prioritization and resource allocation.
\end{abstract}

\maketitle

\section{Introduction}

Technical debt is a metaphor used to describe the consequences of sub-optimal decisions made during software development that prioritize short-term benefits over long-term software maintainability and evolvability \citep{cunningham1992wycash,avgeriou2016managing}.
As technical debt accumulates, it can negatively impact the development process, hindering the ability to make changes to the software, such as fixing bugs or implementing new features.
\textit{Self-Admitted Technical Debt} (SATD) \citep{potdar2014exploratory} is a form of technical debt that is explicitly documented by developers within different software artifacts, such as source code comments, commit messages, issue tracking systems, and pull requests.
For instance, a developer may leave a comment in the code suggesting the removal of unnecessary code in the future: \textit{``TODO: we need to remove the dead code''}.
SATD is, for the most part, complementary to technical debt detected through static code analysis \citep{li2020identification}.
In another example, a developer may document the low performance of a method: \textit{``This method is inefficient and could be refactored for better performance''}.

\myparagraph{The Gap}
In recent years, there has been a growing interest in SATD, with the majority of research focusing on identifying SATD from various sources \citep{ren2019neural,li2022identifying,li2022automatic}.
However, effective SATD management requires not only identification but also estimation of the effort needed for its repayment.
This estimation is essential for prioritizing the repayment of technical debt items and allocating resources efficiently~\citep{li2022self}; sometimes developers are faced with hundreds or even thousands of technical debt items, and prioritizing them becomes a major challenge.
Previous research emphasized the automatic estimation of SATD repayment effort as a critical feature desired by software engineers and managers \citep{li2022self}.
While the measurement of technical debt repayment (commonly known as technical debt \textit{principal}) has been extensively studied and even integrated into industrial tools like SonarQube, this has not been accomplished for SATD. 
This is largely because estimating the repayment effort of SATD involves analyzing natural language documentation by developers, which is fundamentally different from detecting technical debt through methods such as static code analysis.
The current lack of understanding regarding SATD repayment effort, along with the absence of approaches to estimate it, hinders the prioritization of SATD resolution and efficient resource allocation in software development processes.

\myparagraph{Our Approach}
To address this challenge, our research first aims to understand SATD repayment by studying the differences in repayment effort between SATD and non-SATD items, and among different types of SATD items.
After that, we propose an approach for automatically \underline{P}redicting \underline{R}epayment \underline{E}ffort of \underline{S}ATD using \underline{T}extual \underline{I}nformation (PRESTI).
To this end, we gathered all accessible commits, along with their corresponding commit messages and code changes, from 1,060 Apache repositories.
We then identified SATD items by analyzing the commit messages and assessed the repayment effort for each SATD item based on the related code changes.
Using this dataset for training, we assessed the performance of various machine learning models and contrasted the outcomes with a random baseline for automatically estimating the effort required to pay back SATD.
Finally, to further our understanding of SATD repayment, we summarized the keywords that correlate with varying levels of repayment effort, aiming to uncover patterns that could assist in understanding the factors influencing the complexity of repaying SATD.

We decided to use commit messages for SATD identification, instead of other sources (e.g. source code comments, issues), for three main reasons: 1) commit messages typically document resolved SATD, while other sources also discuss SATD that is not resolved yet; 2) commits consist of code changes and are accompanied by commit messages that provide context and purpose for those changes, unlike other sources where code changes and comments are not directly connected; 3) a sufficient quantity of commit messages is available for analysis.

\myparagraph{Contributions}
The primary contributions of this paper are:

\begin{itemize}
    \item \textbf{Assembling an extensive dataset on SATD repayment effort.}
    We gathered a comprehensive dataset consisting of 341,740 SATD items from 2,568,728 commits derived from 1,060 Apache repositories.
    For each SATD item, this dataset includes lines of code added and deleted, the number of files added, modified, and deleted, as well as the significance levels of code changes \citep{fluri2006classifying,fluri2007change}.
    To foster further research in this domain, we make our dataset publicly available \citep{replication}.

    \item \textbf{Presenting the difference in repayment effort between diverse SATD and non-SATD items.}
    Our findings reveal that code/design debt, requirement debt, and test debt necessitate greater repayment effort compared to non-SATD items, whereas documentation debt demands less repayment effort.

    \item \textbf{Introducing and evaluating approaches for automatically estimating SATD repayment effort.}
    Our results demonstrate that SATD effort can be accurately predicted. Particularly, deep learning methods, such as BERT and TextCNN, outperform classic machine learning techniques and the random baseline by a considerable margin in estimating repayment effort.

    \item \textbf{Summarizing keywords associated with various levels of repayment effort.}
    We analyze and summarize the keywords correlated with varying levels of repayment effort that arise during the process of repaying SATD.
    This contribution helps in understanding the factors influencing the complexity of SATD repayment.
\end{itemize}

The organization of this paper is as follows.
The case study design is elaborated in \Cref{sec:approach}.
The results are presented and discussed in \Cref{sec:results} and \Cref{sec:discussion}, respectively.
In \Cref{sec:validity}, we evaluate the threats to validity.
In \Cref{sec:related}, we discuss related work.
Finally, we draw conclusions in \Cref{sec:conclusion}.

\section{Study Design}
\label{sec:approach}

The goal of this study, formulated according to the Goal-Question-Metric \citep{Solingen:02} template is to ``\textit{\textbf{analyze} source code and commit messages \textbf{for the purpose of} investigating the effort required to repay different types of SATD and non-SATD items, automatically estimating this repayment effort, and identifying keywords associated with varying levels of repayment effort \textbf{from the point of view of} software engineers \textbf{in the context of} open-source software.}''
This goal is refined into four research questions (RQs):

\begin{itemize}
    \item \textbf{RQ1:} \textit{Are there differences in repayment effort between SATD and non-SATD items?}\\
    \textbf{Rationale:}
    Prior work \citep{li2020identification,li2022self} shows that resolving SATD items generally takes longer than non-SATD items, but it remains unclear whether this is due to more complex code changes or simply lower priority.
    Understanding the nature of this difference is essential for teams to make informed decisions about resource allocation and to set realistic expectations when planning debt repayment.

    \item \textbf{RQ2:} \textit{Are there differences in repayment effort among different types of SATD items?}\\
    \textbf{Rationale:}
    SATD encompasses several types (e.g., code, design, test, documentation, and requirement debt), each of which may entail different levels of complexity to resolve.
    Characterizing these differences helps developers prioritize which types of SATD to address first, for example by tackling lower-effort types during regular sprints and reserving dedicated resources for higher-effort types.

    \item \textbf{RQ3:} \textit{Can we accurately predict the effort required for SATD repayment based on the SATD text?}\\
    \textbf{Rationale:}
    Estimating SATD repayment effort from textual descriptions remains an open challenge, with no existing automated approaches.
    We explore machine learning techniques to predict the required code changes based on SATD text, which would enable practitioners to prioritize items more effectively and make better-informed decisions about which debt to tackle first.

    \item \textbf{RQ4:} \textit{What keywords are associated with varying levels of repayment effort when repaying SATD?}\\
    \textbf{Rationale:}
    Identifying keywords correlated with different effort levels can provide insights into the complexity behind SATD items. Such keywords can inform targeted guidelines for prioritization, e.g., flagging low-effort items for quick resolution and high-effort items for dedicated planning.
\end{itemize}

\begin{figure}[t]
  \centering
  \includegraphics[width=\linewidth]{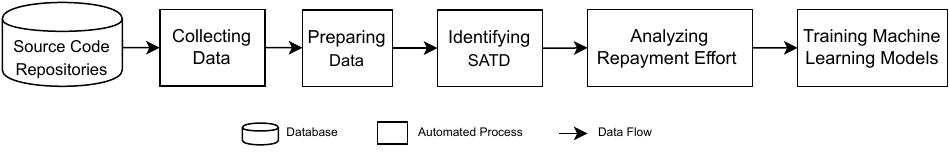}
  \caption{An Overview of PRESTI.}
  \label{f:framework}
  \vspace{-2mm}
\end{figure}

To address our research questions, we use the PRESTI approach outlined in \cref{f:framework}.
First, we \emph{collect data} in terms of commit messages and their corresponding code changes from repositories.
An example\footnote{\url{https://github.com/apache/ant/commit/ecf83d2}} of a commit message along with its associated code changes is presented in \cref{tb:example_result}.
Next, we \emph{prepare data} to ensure consistency and eliminate irrelevant information.
Subsequently, we \emph{identify SATD} items using the commit messages and \emph{analyze the repayment effort} based on the associated code changes.
Finally, we \emph{train machine learning models}.
The subsequent subsections provide detailed information about each of these steps.

\begin{table}[htb]
\caption{Example of a commit message and corresponding code changes}
\label{tb:example_result}
\centering
\resizebox{0.85\columnwidth}{!}{
\begin{tabular}{|l|}
\hline
Commit message: \textbf{implement TODO of configurable buffer size.} \\
\hline
{\code{@@ -131,6 +133,27 @@}} \\
{\code{+ public synchronized void setBufferSize(int bufferSize) \{}} \\
{\code{+\hspace{0.5cm}if (started) \{}} \\
{\code{+\hspace{1cm}throw new IllegalStateException(}} \\
{\code{+\hspace{1.5cm}"Cannot set buffer size on a running StreamPumper");}} \\
{\code{+\hspace{0.5cm}\}}} \\
{\code{+\hspace{0.5cm}this.bufferSize = bufferSize;}} \\
{\code{+ \}}} \\
\hline
\end{tabular}
}
\vspace{-2mm}
\end{table}

\subsection{Collecting Data}

To address our research questions, we focused on analyzing Apache Java projects, owing to their high quality, widespread usage, mature community support, and public availability for access and use.
To gather the necessary data, we obtained all available Apache Java repositories\footnote{\url{https://github.com/orgs/apache/repositories}}.
Our search resulted in a total of 1,060 Apache Java repositories, all of which were included in our study. 
The collection process was carried out using an automated script to ensure consistency and accuracy in the selection of repositories. 
The script, along with the collected dataset, is made available in the replication package \citep{replication} to facilitate future research and validation of our findings.
The collected data includes all available commits and their associated commit messages and code changes from each of the 1,060 Apache Java repositories.

\subsection{Preparing Data}

The data collected from the 1,060 Apache repositories resulted in 2,568,728 commits.
First, we removed merge and rollback commits from the dataset, consistently with previous similar studies \citep{liu2019generating}, as these commits do not provide new information and tend to be substantially larger than other types of commits.
Merge commits represent a point where two branches of code are combined, while rollback commits are used to undo changes made to a codebase.
In the subsequent phase, we removed all commits containing non-English characters.
This was necessary because the tool we used for SATD identification does not support non-English text.
After these steps, we were left with a dataset of 2,382,877 commits for further analysis.

\subsection{Identifying SATD}

To identify SATD items in commit messages, we employed the state-of-the-art approach proposed by Li \textit{et al.} \citep{li2022automatic}.
This approach supports various software artifacts, including commit messages, so it fits with our study goal.
Specifically, this approach uses a CNN and multitask learning technique to identify four types of SATD, namely code/design debt, requirement debt, documentation debt, and test debt, from four software artifacts, which are briefly explained below:

\begin{enumerate}
    \item \textbf{Code/Design Debt:} This type of debt arises from suboptimal decisions made during the coding or design phase of software development. It may include issues such as low-quality code, dead code, or the use of outdated design patterns.

    \item \textbf{Requirement Debt:} Requirement debt refers to situations such as requirements that are only partially implemented or requirements that are implemented but do not fully satisfy all the non-functional requirements.

    \item \textbf{Documentation Debt:} Documentation debt refers to the lack of adequate documentation or low-quality documentation for a software project, including user manuals, developer guides, and code comments.

    \item \textbf{Test Debt:} Test debt is the result of inadequate testing practices, such as insufficient test coverage, poorly designed test cases, or the absence of automated testing.
\end{enumerate}

\begin{table}[ht]
\caption{The number and percentage of different types of SATD items identified in commits.}
\label{tb:satd_statistics}
\begin{center}
\resizebox{\columnwidth}{!}{
\def\arraystretch{1.2}
\begin{tabular}{lcccc}
\hline
Type & Number of SATD items & Percentage of SATD \\
\hline
Code/Design Debt & 260,838 & 76.3 \\
Documentation Debt & 59,175 & 17.3 \\
Requirement Debt & 8,741 & 2.6 \\
Test Debt & 12,986 & 3.8 \\
\hline
\end{tabular}
}
\end{center}
\end{table}

\subsection{Analyzing Repayment Effort}

In this section, we analyze the code changes associated with commit messages that indicate SATD to determine the repayment effort of the SATD items. 
Additionally, we examine the effort of non-SATD code changes using the same methodology as SATD code changes to compare the differences in effort required to address SATD-related vs. non-SATD-related code changes.
Specifically, we assess the effort of code changes from two distinct perspectives: a) the effort required to address SATD items per se, and b) the effort involved in handling ripple effects of the SATD items on the rest of the system.
These two perspectives are essential to understanding the full implications of repaying technical debt, as they capture both the direct and indirect consequences of addressing SATD items.
The first perspective has been the primary focus of existing literature on SATD repayment \citep{mensah2018value,wehaibi2016examining}.
The second perspective is equally important, as addressing technical debt often results in unforeseen consequences on other parts of the software \citep{brown2010managing}.
To capture these two aspects, we employ a comprehensive set of nine different metrics, as detailed below:

\begin{enumerate}[label=\alph*)]
    \item \textbf{Effort Required to Directly Address SATD Items}:
    Inspired by the work of Wehaibi \textit{et al.} \citep{wehaibi2016examining}, we employ five metrics to provide a quantitative assessment of the effort required to address SATD items.
    These metrics include: 1) lines of code added (LA): the number of new lines of code added during changes; 2) lines of code deleted (LD): the number of lines of code removed during changes; 3) file added (FA): the number of new files created and added during changes; 4) file modified (FM): the number of files changed; 5) file deleted (FD): the number of files removed during changes.
    These metrics are commonly employed for cost estimation \citep{nguyen2007sloc,herraiz2006comparison}. 
    For instance, a substantial increase in the lines of code typically suggests a more intricate or significant modification to the software system.
    We employ JGit\footnote{\url{https://www.eclipse.org/jgit}} to analyze repositories and calculate these five metrics.

    \item \textbf{Effort Involved in Handling Ripple Effects}:
    We use four metrics inspired by Fluri \textit{et al.}'s work on the significance level of code changes \citep{fluri2006classifying, fluri2007change} to assess the effort involved in managing ripple effects within the system.
    The significance level expresses how strongly a change may impact other source code entities \citep{fluri2007change}.
    To measure this, we adopt the significance level classification tool\footnote{\url{https://bitbucket.org/sealuzh/tools-changedistiller}} provided by Fluri \textit{et al.} \citep{fluri2006classifying,fluri2007change}, which is based on tree edit operations performed on the abstract syntax tree.
    These metrics include: 1) low significance level code changes (LCC): the number of changes with a low impact on the functionality or performance of the software system; 2) medium significance level code changes (MCC): the number of changes with a moderate impact on the functionality or performance of the software system; 3) high significance level code changes (HCC): the number of changes with a significant impact on the functionality or performance of the software system; 4) crucial significance level code changes (CCC): the number of changes that are critical or essential to the functionality or performance of the software system.
    For instance, code changes in a method body are generally assigned LCC or MCC, while alterations to a class interface are typically designated as HCC or CCC.

\end{enumerate}

\subsection{Training Machine Learning Models}

\subsubsection{Machine Learning Models and Baseline:}

To predict the effort required for SATD repayment based on the SATD text (RQ3), we select appropriate machine learning models.
Although no existing approach specifically targets predicting the repayment effort of SATD, TextCNN and BERT have demonstrated effectiveness in feature extraction for SATD studies \citep{ren2019neural,li2022automatic,li2023automatically}.
Therefore, we choose TextCNN and BERT as our primary machine learning models for predicting the repayment effort of SATD.
We evaluate their performance by comparing them to several classic machine learning approaches and one random baseline.
The methods used in this study are detailed below:

\begin{itemize}
    \item \textbf{TextCNN}:
    TextCNN (Text Convolutional Neural Network) is an effective text classification algorithm proposed by Kim \citep{kim2014convolutional}.
    By employing convolutional layers, TextCNN captures local features and patterns within text data.
    Its ability to efficiently learn discriminative features from text has led to its adoption in numerous SATD identification studies \citep{ren2019neural,li2022identifying,li2022automatic}.
    In this study, we use optimal hyperparameters identified in previous work \citep{li2022automatic} to maximize model performance.
    Specifically, we set the word-embedding dimension at 300, apply filters with five different window sizes (1,2,3,4,5), and use 200 filters for each window size.

    \item \textbf{BERT}:
    BERT (Bidirectional Encoder Representations from Transformers) is a pre-trained transformer-based language model developed by Devlin \textit{et al.} \citep{devlin2018bert} that has shown state-of-the-art performance in various natural language processing tasks, including text classification.
    Given its success in capturing contextual information from text and its proven effectiveness in handling complex language understanding tasks, BERT is well-suited for predicting the repayment effort of SATD based on textual information.
    In this study, we fine-tune BERT base models to leverage their deep understanding of language and context to accurately predict the repayment effort based on the SATD textual information.

    \item \textbf{Classic Machine Learning Approaches (LR, RF, SVR)}:
    Since predicting repayment effort is a regression task, we compare deep learning approaches with Linear Regression (LR), Random Forest (RF), and Support Vector Regression (SVR).
    The input data is vectorized using TF-IDF, and the models are trained using Sklearn implementation with default settings.
    
    \item \textbf{Random Baseline}:
    To provide a benchmark for the performance of our models, we include a random baseline.
    This is necessitated by the absence of any existing approach to estimate the repayment effort of SATD, compelling us to create our own baseline.
    Our objective is to set a reasonable baseline, hence we refrain from using a simple random baseline as used in previous studies \citep{da2017using}.
    Instead, we construct a random baseline that generates predictions based on the mean and standard deviation of the training dataset distribution. 
    This is a common approach in machine learning for establishing a performance benchmark.
\end{itemize}

We implement machine learning models using the PyTorch library and train them on NVIDIA Tesla A100 GPUs.
Our choice of deep learning approaches allows us to effectively extract underlying information and make predictions from unprocessed text, thus eliminating the need for text cleaning.
For evaluation, we shuffle and split the dataset into training, validation, and test sets, adhering to a standard 80/10/10 split.

\subsubsection{Evaluation Metrics}

In this study, we measure the performance of our machine learning models using the Root Mean Square Error (RMSE) metric.
RMSE is a widely adopted metric in regression analysis, employed to assess the accuracy of predicted values in comparison to the actual values.
The RMSE is calculated by taking the square root of the average of the squared differences between the predicted and actual values: $\sqrt{\frac{1}{n} \sum_{i=1}^{n}(actual_i - predicted_i)^2}$, where $n$ is the number of data points, $actual_i$ is the actual value of the $i$th data point, and $predicted_i$ is the predicted value of the $i$th data point.
The lower the RMSE value, the better the performance of the model, indicating a smaller gap between the predicted and ground truth values.

\subsubsection{Keyword Extraction}
\label{sec:keyword}

To address RQ4, we employ a keyword extraction approach to identify keywords associated with varying levels of repayment effort.
This approach is based on the work of Ren \textit{et al.} \citep{ren2019neural}, which uses the backtracking technique on TextCNN to extract n-gram keywords.
First, we feed the text data to the CNN model, and the most important features are selected based on their weights. Next, the corresponding filters are located by backtracking the selected features.
Finally, we locate the n-gram keywords in the input text based on the filter position information.
Specifically, we use the CNN models trained in RQ3 to extract and summarize unigram to five-gram SATD keywords.

\section{Results}
\label{sec:results}

\subsection{RQ1: Are there differences in repayment effort between SATD and non-SATD items?}

To assess the effort needed for the direct resolution of SATD, we provide the mean, median, and trimmed mean values of lines added (LA), lines deleted (LD), and total lines changed in \mbox{\cref{tb:rq1_loc}}.
Because the mean is sensitive to outliers, it provides insight into the overall impact of extreme values in the dataset.
The median is a robust measure that focuses on the central tendency and is resistant to skewness and outliers, making it more appropriate when distributions are highly asymmetric.
The trimmed mean (using a 10\% trimming) balances between the robustness of the median and the efficiency of the mean by removing extreme values from both ends of the data, providing a more stable and reliable estimate of central tendency for skewed distributions \mbox{\citep{kitchenham2017robust,schmoch2020mean}}.
The p-values and effect sizes for comparing SATD and non-SATD are presented in \cref{tb:rq1_loc_p}.
The highest values are highlighted in bold, and the lowest values are underlined for easier comparison.
Although the total lines changed for SATD items are marginally greater than non-SATD items (but not significant according to the p-value), a significant difference exists in the lines added (LA) and lines deleted (LD) for SATD and non-SATD items. 
In particular, SATD changes exhibit lower lines added (LA) and higher lines deleted (LD) compared to non-SATD changes for both mean and median values.

\begin{table}[htb]
\caption{The mean, median, and trimmed mean of lines added (LA) and deleted (LD).}
\label{tb:rq1_loc}
\begin{center}
\resizebox{0.7\columnwidth}{!}{
\def\arraystretch{1.2}
\begin{tabular}{llccccc}
\hline
Group & Measure & LA & LD & Total \\
\hline
SATD & \multirow{2}{*}{Mean} & \underline{68.7} & \textbf{62.8} & \textbf{131.5} \\
Non-SATD & & \textbf{91.0} & \underline{41.7} & \underline{132.7} \\
\hline
SATD & \multirow{2}{*}{Median} & \underline{8.0} & \textbf{6.0} & 14.0 \\
Non-SATD && \textbf{11.0} & \underline{3.0} & 14.0 \\
\hline
SATD & \multirow{2}{*}{Trimmed Mean} & \underline{23.9} & \textbf{18.3} & \textbf{50.6} \\
Non-SATD && \textbf{32.4} & \underline{9.4} & \underline{49.2} \\
\hline
\end{tabular}
}
\end{center}
\caption{P-values and effect sizes for comparing SATD and non-SATD.}
\label{tb:rq1_loc_p}
\begin{center}
\resizebox{0.5\columnwidth}{!}{
\def\arraystretch{1.2}
\begin{tabular}{lccccc}
\hline
Test & LA & LD & Total \\
\hline
P-value & 0.00 & 0.00 & 0.10 \\
Effect Size & -0.09 & 0.12 & 0.00 \\
\hline
\end{tabular}
}
\end{center}
\vspace{-3mm}
\end{table}

\begin{table}[htb]
\caption{The mean, median, and trimmed mean of file added (FA), deleted (FD), and modified (FM).}
\label{tb:rq1_file}
\begin{center}
\resizebox{0.7\columnwidth}{!}{
\def\arraystretch{1.2}
\begin{tabular}{llccccc}
\hline
Group & Measure & FA & FD & FM & Total \\
\hline
SATD & \multirow{2}{*}{Mean} & \underline{0.34} & \textbf{0.32} & \underline{3.64} & \underline{4.31} \\
Non-SATD && \textbf{0.62} & \underline{0.21} & \textbf{3.92} & \textbf{4.77} \\
\hline
SATD & \multirow{2}{*}{Median} & 0.00 & 0.00 & 1.00 & 1.00 \\
Non-SATD && 0.00 & 0.00 & 1.00 & 1.00 \\
\hline
SATD & \multirow{2}{*}{Trimmed Mean} & \underline{0.00} & 0.00 & \textbf{1.92} & 2.24 \\
Non-SATD && \textbf{0.09} & 0.00 & \underline{1.83} & 2.24 \\
\hline
\end{tabular}
}
\end{center}
\caption{P-values and effect sizes for comparing SATD and non-SATD.}
\label{tb:rq1_file_p}
\begin{center}
\resizebox{0.6\columnwidth}{!}{
\def\arraystretch{1.2}
\begin{tabular}{lccccc}
\hline
Test & FA & FD & FM & Total \\
\hline
P-value & 0.00 & 3.23e-54 & 7.42e-23 & 1.46e-48 \\
Effect Size & -0.07 & 0.03 & -0.01 & -0.02 \\
\hline
\end{tabular}
}
\end{center}
\end{table}

For assessing the effort in direct resolution of SATD in terms of file changes, we display the results for file added (FA), file deleted (FD), file modified (FM), and the total number of files changed in \cref{tb:rq1_file}.
It can be observed that the total number of files changed for SATD repayment is marginally lower than non-SATD changes, but with a small effect size (-0.02).
Furthermore, the results show that SATD changes have a lower number of file added (FA) compared to non-SATD changes, while the differences in file deleted (FD) and file modified (FM) are relatively minor.

\begin{table}[htb]
\caption{The mean, median, and trimmed mean of low (LCC), medium (MCC), high (HCC), and crucial (CCC) significance level code changes.}
\label{tb:rq1_impact}
\begin{center}
\resizebox{0.8\columnwidth}{!}{
\def\arraystretch{1.2}
\begin{tabular}{llccccc}
\hline
Group & Measure & LCC & MCC & HCC & CCC & Total \\
\hline
SATD & \multirow{2}{*}{Mean} & \textbf{6.22} & \textbf{4.33} & \textbf{0.66} & \textbf{0.77} & \textbf{11.98} \\
Non-SATD && \underline{5.63} & \underline{3.10} & \underline{0.39} & \underline{0.50} & \underline{9.61} \\
\hline
SATD & \multirow{2}{*}{Median} & 0.00 & 0.00 & 0.00 & 0.00 & 0.00 \\
Non-SATD && 0.00 & 0.00 & 0.00 & 0.00 & 0.00 \\
\hline
SATD & \multirow{2}{*}{Trimmed Mean} & \textbf{1.90} & \textbf{0.94} & \textbf{0.05} & \textbf{0.04} & \textbf{3.81} \\
Non-SATD && \underline{1.76} & \underline{0.56} & \underline{0.00} & \underline{0.00} & \underline{2.92} \\
\hline
\end{tabular}
}
\end{center}
\caption{P-values and effect sizes for comparing SATD and non-SATD.}
\label{tb:rq1_impact_p}
\begin{center}
\resizebox{0.8\columnwidth}{!}{
\def\arraystretch{1.2}
\begin{tabular}{lccccc}
\hline
Test & LCC & MCC & HCC & CCC & Total \\
\hline
P-value & 4.18e-36 & 5.57e-237 & 3.06e-201 & 1.17e-138 & 1.35e-173 \\
Effect Size & 0.02 & 0.08 & 0.08 & 0.06 & 0.06 \\
\hline
\end{tabular}
}
\end{center}
\end{table}

To evaluate the effort involved in handling ripple effects, \cref{tb:rq1_impact} presents the mean, median, and trimmed mean values of low (LCC), medium (MCC), high (HCC), and crucial (CCC) significance level code changes.
Also, the p-values and effect sizes between SATD and non-SATD are presented in \cref{tb:rq1_impact_p}.
The results reveal a significantly higher number of code changes across all significance levels for SATD repayment compared to non-SATD changes.
Notably, SATD repayment changes involve a substantially higher number of medium-significance code changes (MCC) and high-significance code changes (HCC) compared to non-SATD changes with an effect size of 0.08.
Our statistical analysis confirms that the differences between SATD and non-SATD changes are statistically significant with respect to LA, LD, FA, FD, FM, and different levels of code changes, with a p-value less than 0.05, using the Mann-Whitney test \citep{mann1947test}.

\takeaway{
\textbf{RQ1:} Although SATD and non-SATD items require \textbf{similar} levels of effort for direct resolution (mean total lines changed: \textbf{131.5} vs. \textbf{132.7}), SATD items demand \textbf{higher} effort in managing the associated ripple effects, with significantly more medium- and high-significance code changes (effect size = \textbf{0.08}, p $<$ 0.05).
}

\subsection{RQ2: Are there differences in repayment effort among different types of SATD items?}
\label{sec:rq2}

We evaluate the direct resolution effort for various SATD types by presenting the mean and median values of lines added (LA), lines deleted (LD), and total lines changed for each SATD type and non-SATD items in \cref{tb:rq2_loc} and \cref{tb:rq2_loc_median}. 
To highlight the differences between distinct types of SATD and non-SATD, we report the rank obtained from the ScottKnottESD test \citep{tantithamthavorn2018impact} in \cref{tb:rq2_loc} and \cref{tb:rq2_loc_median} within parentheses. 
The Scott-Knott Effect Size Difference (ESD) test is a multiple comparison approach that employs hierarchical clustering to partition the set of treatment averages into statistically distinct groups with non-negligible differences \citep{tantithamthavorn2018impact}. 
The Scott-Knott ESD test generates the ranking while ensuring that (1) the magnitude of the difference for all treatments in each group is negligible, and (2) the magnitude of the difference of treatments between groups is non-negligible.

\begin{table}[ht]
\caption{The average number of lines added (LA) and deleted (LD).}
\label{tb:rq2_loc}
\begin{center}
\resizebox{0.8\columnwidth}{!}{
\def\arraystretch{1.2}
\begin{tabular}{lccc}
\hline
Type & LA & LD & Total \\
\hline
Code/Design Debt & 69.1(3) & \textbf{73.1(1)} & 142.2(2) \\
Documentation Debt & \underline{48.5(3)} & \underline{24.2(3)} & \underline{72.7(3)} \\
Requirement Debt & \textbf{126.0(1)} & 43.7(2) & \textbf{169.7(1)} \\
Test Debt & 118.2(2) & 47.9(2) & 166.1(2) \\
\hline
Non-SATD & 91.0(2) & 41.7(2) & 132.7(2) \\
\hline
\end{tabular}
}
\end{center}
\caption{The median number of lines added (LA) and deleted (LD).}
\label{tb:rq2_loc_median}
\begin{center}
\resizebox{0.8\columnwidth}{!}{
\def\arraystretch{1.2}
\begin{tabular}{lccc}
\hline
Type & LA & LD & Total \\
\hline
Code/Design Debt & 8.0(3) & \textbf{9.0(1)} & 17.0(2) \\
Documentation Debt & \underline{4.0(3)} & \underline{2.0(3)} & \underline{6.0(3)} \\
Requirement Debt & 15.0(1) & 5.0(2) & 20.0(1) \\
Test Debt & \textbf{25.0(2)} & 4.0(2) & \textbf{29.0(2)} \\
\hline
Non-SATD & 11.0(2) & 3.0(2) & 14.0(2) \\
\hline
\end{tabular}
}
\end{center}
\vspace{-3mm}
\end{table}

\begin{table}[ht]
\caption{The average number of file added (FA), deleted (FD), and modified (FM).}
\label{tb:rq2_file}
\begin{center}
\resizebox{0.9\columnwidth}{!}{
\def\arraystretch{1.2}
\begin{tabular}{lcccc}
\hline
Type & FA & FD & FM & Total \\
\hline
Code/Design Debt & 0.34(2) & \textbf{0.41(1)} & \textbf{3.86(1)} & \textbf{4.61(1)} \\
Documentation Debt & \underline{0.24(2)} & \underline{0.10(1)} & 3.02(1) & \underline{3.36(1)} \\
Requirement Debt & \textbf{0.83(1)} & 0.19(1) & 3.28(1) & 4.30(1) \\
Test Debt & \textbf{0.83(1)} & 0.24(1) & \underline{2.43(1)} & 3.50(1) \\
\hline
Non-SATD & 0.62(2) & 0.21(1) & 3.92(1) & 4.77(1) \\
\hline
\end{tabular}
}
\end{center}
\caption{The median number of file added (FA), deleted (FD), and modified (FM).}
\label{tb:rq2_file_median}
\begin{center}
\resizebox{0.8\columnwidth}{!}{
\def\arraystretch{1.2}
\begin{tabular}{lcccc}
\hline
Type & FA & FD & FM & Total \\
\hline
Code/Design Debt & 0.0(1) & 0.0(1) & 1.0(1) & 1.0(1) \\
Documentation Debt & 0.0(1) & 0.0(1) & 1.0(1) & 1.0(1) \\
Requirement Debt & 0.0(2) & 0.0(1) & 1.0(1) & 1.0(1) \\
Test Debt & 0.0(2) & 0.0(1) & 1.0(1) & 1.0(1) \\
\hline
Non-SATD & 0.0(1) & 0.0(1) & 1.0(1) & 1.0(1) \\
\hline
\end{tabular}
}
\end{center}
\end{table}

As can be seen in \cref{tb:rq2_loc} and \cref{tb:rq2_loc_median}, documentation debt changes demonstrate significantly lower levels of lines added (LA) and lines deleted (LD) compared to non-SATD changes.
In contrast, requirement debt changes exhibit a substantial increase in the number of lines added (LA) compared to non-SATD changes.
Code/design debt changes show a lower number of lines added (LA) and a higher number of lines deleted (LD) compared to non-SATD changes.

\begin{table}[ht]
\caption{The average number of low (LCC), medium (MCC), high (HCC), and crucial (CCC) significance level code changes.}
\label{tb:rq2_impact}
\begin{center}
\resizebox{\columnwidth}{!}{
\def\arraystretch{1.2}
\begin{tabular}{lccccc}
\hline
Type & LCC & MCC & HCC & CCC & Total \\
\hline
Code/Design Debt & 7.19(2) & 5.13(1) & 0.79(1) & 0.92(1) & 14.03(1) \\
Documentation Debt & \underline{1.28(3)} & \underline{0.66(3)} & \underline{0.14(1)} & \underline{0.12(1)} & \underline{2.20(3)} \\
Requirement Debt & \textbf{9.76(1)} & \textbf{5.83(1)} & \textbf{0.87(1)} & \textbf{1.02(1)} & \textbf{17.48(1)} \\
Test Debt & 8.51(1) & 4.08(2) & 0.49(1) & 0.56(1) & 13.64(1) \\
\hline
Non-SATD & 5.63(3) & 3.10(3) & 0.39(1) & 0.50(1) & 9.61(2) \\
\hline
\end{tabular}
}
\end{center}
\caption{The median number of low (LCC), medium (MCC), high (HCC), and crucial (CCC) significance level code changes.}
\label{tb:rq2_impact_median}
\begin{center}
\resizebox{\columnwidth}{!}{
\def\arraystretch{1.2}
\begin{tabular}{lccccc}
\hline
Type & LCC & MCC & HCC & CCC & Total \\
\hline
Code/Design Debt & \underline{0.00(2)} & 0.00(1) & 0.00(1) & 0.00(1) & \underline{0.00(1)} \\
Documentation Debt & \underline{0.00(3)} & 0.00(3) & 0.00(1) & 0.00(1) & \underline{0.00(3)} \\
Requirement Debt & 1.00(1) & 0.00(1) & 0.00(1) & 0.00(1) & 1.00(1) \\
Test Debt & \textbf{2.00(1)} & 0.00(2) & 0.00(1) & 0.00(1) & \textbf{2.00(1)} \\
\hline
Non-SATD & 0.00(3) & 0.00(3) & 0.00(1) & 0.00(1) & 0.00(2) \\
\hline
\end{tabular}
}
\end{center}
\vspace{-3mm}
\end{table}

\Cref{tb:rq2_file} and \cref{tb:rq2_file_median} summarize the results of file added (FA), file deleted (FD), file modified (FM), and the total number of files changed for each SATD type and non-SATD items, further assessing the direct resolution effort.
We can observe that documentation debt changes result in fewer file updates, whereas changes associated with requirement and test debt lead to a substantial increase in the number of file added compared to non-SATD changes; this is similar to the results in \cref{tb:rq2_loc}.
Moreover, we observe a similar trend for code/design debt changes, with fewer file added (FA) and more file deleted (FD) than non-SATD changes. 
Interestingly, our findings indicate that test debt changes have the least number of file modified (FM) compared to other types of SATD and non-SATD changes.

To evaluate the effort involved in handling ripple effects, \Cref{tb:rq2_impact} and \cref{tb:rq2_impact_median} summarize the number of code changes of low (LCC), medium (MCC), high (HCC), and crucial (CCC) significance levels, for each type of SATD and non-SATD changes.
As we can see, documentation debt changes result in the lowest number of code changes across different significance levels; especially for LCC and MCC, the differences are significant.
In contrast, SATD changes associated with requirement debt exhibit the highest number of code changes across all significance levels, significantly exceeding non-SATD changes.
Code/design and test debt changes display a similar trend, following requirement debt in terms of the number of code changes across different significance levels, which are also significantly higher than non-SATD changes.

\takeaway{
\textbf{RQ2:} Various types of SATD items demonstrate \textbf{distinct levels} of repayment effort. \textbf{Documentation debt} presents the lowest repayment effort (mean total lines: \textbf{72.7}), even lower than non-SATD items (\textbf{132.7}), while \textbf{requirement debt} leads to the highest (\textbf{169.7}) among all SATD types.
}

\subsection{RQ3: Can we accurately predict the effort required for SATD repayment based on the SATD text?}
\label{sec:rq3}

To assess the efficacy of various machine learning approaches in predicting the required code changes to repay SATD, we compared two deep learning methods (i.e., BERT and TextCNN) and three classical machine learning methods (i.e., RF, LR, and SVR) against a baseline (i.e., random).
Due to the substantial variability in the predicted values, we implemented a logarithmic transformation on the target variables, such as LA and FA, to stabilize the variance and normalize the distribution.
This transformation produced a more normal distribution, which subsequently improved the learning capabilities of the evaluated machine learning models.
Once we obtain the results, we convert them back to their original values for evaluation.

\begin{table}[htb]
\caption{Comparison of RMSE performance: machine learning models versus baseline approach for predicting lines added (LA), lines deleted (LD), and total lines of code changed (LT).}
\label{tb:rq3_1}
\begin{center}
\resizebox{0.7\columnwidth}{!}{
\def\arraystretch{1.2}
\begin{tabular}{lccccc}
\hline
Approach & LA & LD & LT & Average \\
\hline
BERT\textsubscript{BASE} & 11.1(5) & \textbf{10.0(6)} & \textbf{19.5(6)} & \textbf{13.5(5)} \\
TextCNN & \textbf{10.8(6)} & 10.3(5) & 19.8 (5)& 13.6(5) \\
RF & 15.5(3) & 14.5(3) & 26.3(3) & 18.8(3) \\
LR & 20.1(2) & 18.5(2) & 34.7(2) & 24.4(2) \\
SVR & 12.8(4) & 11.9(4) & 23.4(4) & 16.0(4) \\
\hline
Random & 32.8(1) & 27.5(1) & 54.0(1) & 38.1(1) \\
\hline
\end{tabular}
}
\end{center}
\end{table}

\begin{table}[htb]
\caption{Comparison of RMSE performance: machine learning models versus baseline approach for predicting file added, deleted, modified files, and the total number of files affected.}
\label{tb:rq3_3}
\begin{center}
\resizebox{0.9\columnwidth}{!}{
\def\arraystretch{1.2}
\begin{tabular}{lcccccc}
\hline
Approach & FA & FD & FM & FT & Average \\
\hline
BERT\textsubscript{BASE} & \textbf{0.33(5)} & \textbf{0.30(5)} & \textbf{2.03(5)} & \textbf{2.30(6)} & \textbf{1.24(5)} \\
TextCNN & \textbf{0.33(5)} & \textbf{0.30(5)} & 2.05(5) & 2.32(5) & 1.25(5) \\
RF & 0.42(3) & 0.38(3) & 2.44(3) & 2.79(3) & 1.51(3) \\
LR & 0.53(2) & 0.51(2) & 2.86(2) & 3.25(2) & 1.79(2) \\
SVR & 0.38(4) & 0.36(4) & 2.21(4) & 2.50(4) & 1.36(4) \\
\hline
Random & 0.63(1) & 0.59(1) & 3.36(1) & 3.84(1) & 2.10(1) \\
\hline
\end{tabular}
}
\end{center}
\end{table}

The results presented in \Cref{tb:rq3_1} provide a comprehensive comparison of different approaches in predicting the lines added (LA), lines deleted (LD), and total lines changed (LT) required for SATD repayment based on the SATD text.
It is noted that the best results are highlighted in bold.
To show the differences between different approaches, we also report the rank obtained from the ScottKnottESD test \citep{tantithamthavorn2018impact} in \cref{tb:rq3_1} within parentheses. 
As shown in \Cref{tb:rq3_1}, the BERT-based approach achieves the lowest RMSE values for LD (10.0) and LT (19.5), while TextCNN achieves the lowest RMSE for LA (10.8). Overall, BERT attains the best average RMSE of 13.5, closely followed by TextCNN with an average RMSE of 13.6.
The BERT-based approach significantly improves the average RMSE from 38.1 (baseline method) to 13.5.
In comparison, the classical machine learning approaches yield average RMSE scores ranging from 16.0 to 24.4.
Although these scores are significantly lower than the random baseline (38.1), they are still higher than the scores achieved by the BERT approach.

We also predicted the number of files that need to be added (FA), deleted (FD), and modified (FM), as well as the total number of affected files (FT), and showed the results in \Cref{tb:rq3_3}.
The BERT-based approach consistently demonstrates the lowest RMSE across all predicted values, with values of 0.33 for FA, 0.30 for FD, 2.03 for FM, and 2.30 for FT, resulting in an average RMSE of 1.24.
Similarly, the performance of TextCNN is slightly worse than BERT (average 1.25 vs 1.24).
Moreover, the average RMSE scores obtained by the classical machine learning approaches range from 1.36 to 1.79, which are substantially lower than the baseline (2.10).

\begin{table}[htb]
\caption{Comparison of RMSE performance: machine learning models versus baseline approach for predicting different levels of code changes.}
\label{tb:rq3_2}
\begin{center}
\resizebox{0.9\columnwidth}{!}{
\def\arraystretch{1.2}
\begin{tabular}{lcccccc}
\hline
Approach & LCC & MCC & HCC & CCC & Average \\
\hline
BERT\textsubscript{BASE} & \textbf{2.01(6)} & \textbf{1.58(6)} & 0.51(5) & \textbf{0.51(5)} & \textbf{1.15(6)} \\
TextCNN & 2.07(5) & 1.66(5) & \textbf{0.47(6)} & \textbf{0.51(5)} & 1.18(5) \\
RF & 2.92(3) & 2.09(3) & 0.63(3) & 0.64(3) & 1.57(3) \\
LR & 3.46(2) & 2.54(2) & 0.77(2) & 0.79(2) & 1.89(2) \\
SVR & 2.37(4) & 1.70(4) & 0.55(4) & 0.55(4) & 1.29(4) \\
\hline
Random & 5.01(1) & 3.48(1) & 0.94(1) & 0.99(1) & 2.60(1) \\
\hline
\end{tabular}
}
\end{center}
\end{table}

In addition to predicting the changed lines of code and number of files required to repay SATD, we investigated the application of machine learning approaches to predict the number of code changes of the various significance levels, based on the SATD text.
\Cref{tb:rq3_2} presents the RMSE performance of the approaches in predicting different significance levels of code changes required for SATD repayment.
As shown in the table, the BERT-based approach still achieved the best performance in predicting different significance levels of code changes, with an average RMSE of 1.15.
Specifically, the model exhibits the best performance in predicting LCC, MCC, HCC, and CCC with RMSE of 2.01, 1.58, 0.51, and 0.51.
Notably, the BERT-based approach is surpassed by TextCNN by a small margin when predicting HCC.
Additionally, the three classical machine learning models exhibit better RMSE performance than the baseline approach (2.60), with values ranging from 1.29 to 1.89.

\takeaway{
\textbf{RQ3:} Machine learning approaches, specifically BERT and TextCNN, can be \textbf{effective in predicting the effort required for SATD repayment} based on the SATD text. BERT achieves the best average RMSE of \textbf{13.5} for lines of code prediction, reducing the baseline error (\textbf{38.1}) by \textbf{64.6\%}.
}

\subsection{RQ4: What keywords are associated with varying levels of repayment effort when repaying SATD?}
\label{sec:rq4}

We begin by summarizing the keywords linked to the effort required for directly addressing SATD items.
Using the deconvolution technique (refer to \cref{sec:keyword}), we identify and present the top keywords associated with low and high lines of code changed (including LA and LD) as well as low and high numbers of files changed (including FA, FD, and FM) in \cref{tb:rq4_1}.
We opted to summarize keywords for lines of code changed and the number of files changed since they are the most prevalent metrics for effort estimation in software engineering \citep{nguyen2007sloc,herraiz2006comparison}.
Unique keywords are highlighted in bold.
Our analysis reveals that when the lines of code and number of file modified are low, the keywords generally relate to typos, error message updates, warning message updates, or code comments.
An example of improving an error message is presented below: 15 lines of code are added, and 6 are removed.

\begin{table}[ht]
\caption{Top keywords associated with low or high levels of SATD repayment effort with respect to the number of lines of code modified and the number of file modified.}
\label{tb:rq4_1}
\begin{center}
\resizebox{0.9\columnwidth}{!}{
\def\arraystretch{1.2}
\begin{tabular}{cccc}
\hline
Low \# Lines & High \# Lines & Low \# Files & High \# Files \\
\hline
typo & code cleanup &  typo & header \\
unused import & \textbf{formatting} &  unused import & \textbf{interface} \\
error message & \textbf{more tests} & comment & code cleanup \\
comment & \textbf{documentation} & \textbf{warning} & \textbf{annotation} \\
\textbf{logging} & \textbf{work in progress} & debug & naming \\
\textbf{javadoc} & \textbf{improvement} & \textbf{workaround} & \textbf{class} \\
\textbf{minor} & rename & \textbf{proper} & \textbf{tidy up} \\
\textbf{update} & \textbf{support for} & \textbf{variable} & \textbf{files} \\
debug & header & error message & \textbf{extension point} \\
\hline
\end{tabular}
}
\end{center}
\vspace{-2mm}
\end{table}

\begin{displayquote}
\textit{``Modify parsing of the HDFS directory to be less restrictive, provide a better error message, fail if directory can't be created, handle slf4j jars.''} - [Accumulo-cd31dc]
\end{displayquote}

We also observe that SATD items involving unused imports, logging, workarounds, or debugging consistently require low repayment effort, measured in terms of lines of code and number of files.
For example, developers removed unused imports by modifying just 1 file and deleting 3 lines of code:

\begin{displayquote}
\textit{``Minor: organized imports. This also removed an unused import that was Java 7.''} - [Stanbol-ee22da]
\end{displayquote}

In contrast, SATD items related to code cleanup (e.g., \emph{code cleanup}, \emph{formatting}, and \emph{rename}), tests, documentation (e.g., \emph{documentation} and \emph{license header}), and requirements (e.g., \emph{work in progress}, \emph{improvement}, and \emph{support for}) generally demand more lines of code to repay.
For instance, a developer cleaned up the Jackrabbit project by adding 84 and removing 85 lines of code:

\begin{displayquote}
\textit{``JCR-3525 code cleanup and java doc comments; remove the option of checking in batches instead of all at once: this broke during a previous refactoring and is not a user friendly option since it is very slow.''} - [Jackrabbit-ebf78b]
\end{displayquote}

Interestingly, keywords for high numbers of file modified differ from those for high lines of code.
While keywords related to high numbers of file modified also pertain to code cleanup (e.g., \emph{code cleanup}, \emph{naming}, and \emph{tidy up}), they involve changes to interfaces and classes as well (e.g., \emph{interface}, \emph{class}, and \emph{extension point}).
In the following example, developers removed an interface by modifying 3 files and deleting 1 file:

\begin{displayquote}
\textit{``JCR-2092: remove old SameNode interface.''} - [Jackrabbit-24f627]
\end{displayquote}

\begin{table}[htb]
\caption{Top keywords that are associated with varying levels of significant changes for repaying SATD.}
\label{tb:rq4_2}
\begin{center}
\resizebox{0.9\columnwidth}{!}{
\def\arraystretch{1.2}
\begin{tabular}{cccc}
\hline
LCC & MCC & HCC & CCC \\
\hline
logging & handling & unused code & unused code \\
\textbf{exception} & logging & interface & interface \\
handling & simplify & API & refactoring \\
\textbf{test} & \textbf{logic} & implementation & API \\
output & \textbf{catch} & code cleanup & support \\
\textbf{cast} & output & refactoring & \textbf{deprecated code} \\
simplify & code cleanup & support & implementation \\
\textbf{findbugs} & leak & \textbf{checkstyle errors} & \textbf{constructor} \\
leak & implementation & \textbf{redundant} & \textbf{endpoints} \\
\hline
\end{tabular}
}
\end{center}
\vspace{-2mm}
\end{table}

To summarize keywords connected to the effort in handling ripple effects, we provide an overview of the keywords identified for code changes with varying levels of significance (i.e., low, medium, high, and crucial) during SATD repayment in \Cref{tb:rq4_2}.
Note that unique keywords (exclusive to one significance level) are highlighted in bold for easy reference.
The keywords show that changes with low or medium significance levels primarily focus on exception handling, logging, tests, logic improvement, and fixing leak issues.
On the other hand, changes with high or crucial significance levels predominantly involve cleaning up unused code, modifying interfaces, refactoring code, and implementing new requirements.
Some keywords are shared between different levels of significance, as certain tasks in software development are fundamental and ubiquitous across all complexity levels.

\takeaway{
\textbf{RQ4:} Different types of SATD repayment efforts are associated with distinct keywords.
\textbf{Low-effort repayments} typically involve typos, error messages, and code comments (e.g., \textbf{1} file modified, \textbf{3} lines deleted), while \textbf{high-effort repayments} often require code cleanup, modifying interfaces, and implementing requirements (e.g., \textbf{84} lines added and \textbf{85} removed).
}

\section{Discussion}
\label{sec:discussion}

\myparagraph{SATD Repayment Resembles Refactoring, Not Feature Development}
Our results show that while SATD and non-SATD items require similar total lines changed (131.5 vs.\ 132.7), the nature of the changes differs markedly. SATD repayment involves fewer lines added and more lines deleted compared to non-SATD changes, a pattern that extends to files as well (lower FA, higher FD). This suggests that SATD repayment is predominantly a refactoring activity (i.e., removing and restructuring problematic code) rather than adding new functionality. This contrasts with the findings of Wehaibi et al.\ \citep{wehaibi2016examining}, who reported significantly higher effort for SATD changes. The discrepancy likely stems from differences in SATD identification methods: our study uses commit messages, while theirs relied on 62 code comment keywords \citep{potdar2014exploratory} that may capture more severe SATD items.

\myparagraph{SATD Repayment Creates Larger Ripple Effects}
Despite similar direct resolution effort, SATD repayment changes involve a substantially higher number of code changes across all significance levels compared to non-SATD changes (effect size = 0.08 for MCC and HCC, p $<$ 0.05). This indicates that addressing SATD may require more extensive and diverse code modifications that impact multiple aspects of the codebase. A possible explanation is that SATD items tend to be more deeply embedded within the code, necessitating broader intervention. For practitioners, this emphasizes the importance of early detection and resolution of SATD to minimize ripple effects.

\myparagraph{SATD Types Should Be Treated Distinctly, Not as a Monolithic Entity}
Our analysis reveals that different SATD types require vastly different levels of repayment effort. Documentation debt demands the lowest effort (mean total lines: 72.7), even less than non-SATD items (132.7), as it often only requires updating comments without altering the codebase. In contrast, requirement debt leads to the highest effort (169.7), as developers need to implement new features or enhance existing ones. Code/design debt exhibits a characteristic refactoring pattern (lower LA, higher LD), while test debt requires substantial additions (median LA: 25.0). These findings suggest that researchers and practitioners should treat each SATD type distinctly when prioritizing repayment, and that future work should extend to types that are difficult to detect automatically, such as architecture or build debt.

\myparagraph{Deep Learning Enables Effective Effort Prediction from Text}
BERT and TextCNN significantly outperform classical machine learning methods in predicting SATD repayment effort from text. BERT achieves the best average RMSE of 13.5 for lines of code prediction, reducing the baseline error (38.1) by 64.6\%. This superior performance can be attributed to the ability of deep learning models to capture complex patterns and semantic relationships within SATD text. By providing estimated repayment effort for each SATD item, these models can help development teams make informed resource allocation decisions. However, the predictions are not yet perfect, and our models are trained exclusively on commit message text. Future research should explore predicting repayment effort across different SATD documentation sources (code comments, issue trackers, pull requests) and investigate hybrid approaches that combine automated predictions with human expertise.

\myparagraph{Keywords Offer an Interpretable Complement to Model Predictions}
Using a deconvolution technique, we identified keywords associated with varying levels of repayment effort. Low-effort keywords (e.g., \emph{typo}, \emph{unused import}, \emph{error message}) relate to simple fixes, while high-effort keywords (e.g., \emph{code cleanup}, \emph{work in progress}, \emph{interface}) correspond to more complex structural changes. These keywords provide an interpretable and lightweight alternative for estimating effort when deep learning predictions are unavailable. Integrating these keywords into automated tools, for example by highlighting high-effort indicators in SATD items, can help developers quickly gauge repayment complexity and prioritize accordingly. Combining keyword-based insights with model predictions can provide a comprehensive understanding of the effort associated with different SATD items.

\section{Threats to Validity}
\label{sec:validity}

\subsection{Construct Validity}

A first concern is that we used lines of code, files changed, and significance levels of code changes as proxies for repayment effort, which may not capture all aspects of the effort involved in SATD repayment.
We mitigate this by employing widely-accepted metrics that have been extensively used in prior software engineering studies to quantify code change effort.
A second concern is the presence of tangled commits, where an SATD fix is bundled together with other changes that are not related to the SATD being fixed, which could inflate the measured effort.
We mitigate this by following the approach of Liu et al.\ \citep{liu2019generating}: we removed merge and rollback commits, as these do not provide new information and tend to be substantially larger than regular commits, thereby reducing the likelihood of including tangled commits in our analysis.
A third concern is that developers may admit the presence of SATD in commit messages without actually fixing it, which would lead to incorrect effort estimates.
We mitigate this by noting that Li et al.\ \citep{li2023automatically} found that only about 10.75\% (1,724 out of 36,037) of commits document unsolved SATD items.
Given this relatively low proportion and the current absence of automated approaches to distinguish solved from unsolved SATD in commit messages, we used the identified SATD directly to study repayment effort.
A fourth concern is that the machine learning model used to identify SATD items \citep{li2023automatically} may miss certain SATD types that are not typically described in commit messages, potentially affecting the completeness of our dataset.
We mitigate this by using a well-established model from prior research that has demonstrated high accuracy in identifying various types of SATD across multiple datasets.

\subsection{Reliability}

A concern regarding reliability is that machine learning models inherently possess randomness due to their training process (e.g., random weight initialization and mini-batch sampling), which may affect the consistency and reproducibility of results.
We mitigate this by reporting the average performance across multiple runs of the models, reducing the impact of any single run's variability.
Moreover, we made our collected dataset and code publicly available in the replication package \citep{replication}, enabling independent replication and verification of the results.

\subsection{External Validity}

A concern regarding external validity is that our study focused exclusively on Java open-source Apache projects, which may not be representative of all software projects, programming languages, or application domains.
We mitigate this by including 1,060 Apache repositories that vary considerably in size and complexity, thereby increasing the diversity of our dataset.
Future studies should investigate SATD repayment effort across different programming languages, domains, and project types, as well as incorporate more diverse sources of data such as issue trackers and code review comments.

\section{Related Work}
\label{sec:related}

This study aims to investigate the effort required to repay SATD.
Accordingly, we categorize the relevant literature into two main sections: A) prior research on SATD in general, and B) earlier studies related to the analysis of repayment effort for technical debt.

\subsection{Self-Admitted Technical Debt in General}

The initial investigation into SATD in source code comments was conducted by Potdar and Shihab \citep{potdar2014exploratory}, who examined four open-source projects and discovered that SATD comments were present in 2.4\% to 31\% of source files, and only 26.3\% to 63.5\% of the identified SATD comments were removed after being introduced.
Maldonado and Shihab \citep{maldonado2015detecting} expanded on this work by classifying SATD into five categories (design, requirement, defect, documentation, and test debt) based on the analysis of 33,000 code comments from five open-source projects.
Their results showed that design debt was the most frequent form of SATD, accounting for 42\% to 84\% of classified cases.
Subsequent to the exploration of SATD, researchers have shown considerable interest in devising automated methods for SATD detection.
Several machine learning approaches \citep{ren2019neural, li2022identifying, li2022automatic} have been employed to detect different types of SATD items from various sources.
Ren \textit{et al.} \citep{ren2019neural} developed a Convolutional Neural Network-based approach to enhance the accuracy and explainability of SATD detection, particularly for cross-project prediction.
Li \textit{et al.} \citep{li2022identifying} generated a dataset of 4,200 issues from seven open-source projects and proposed a machine learning approach to detect SATD in issue tracking systems, outperforming baseline methods, benefiting from knowledge transfer, and extracting intuitive SATD keywords.
Li \textit{et al.} \citep{li2022automatic} also introduced an automated SATD identification approach from multiple sources, such as source code comments, commit messages, pull requests, and issue tracking systems, leveraging a multitask learning technique.

Additionally, researchers explored the removal of SATD.
Maldonado \textit{et al.} \citep{maldonado2017empirical} evaluated the removal of SATD in five open-source projects and observed that the majority of SATD is eventually removed, mostly by those who introduced it.
They found that it takes 18 to 172 days to remove SATD comments on average.
Zampetti \textit{et al.} \citep{zampetti2018self} also studied the removal of SATD in five Java open-source projects and discovered that 20\% to 50\% of SATD is unintentionally removed, and only 8\% of debt removal is documented in commit messages.

\subsection{Repayment Effort for Technical Debt}

Numerous studies have investigated the effort of paying back technical debt. 
Xiao \textit{et al.} \citep{xiao2016identifying} introduced a novel approach for the automatic detection, quantification, and modeling of architectural debt in software systems.
The authors quantified effort by measuring the number of lines of code modified and committed to fix bugs. 
Their evaluation carried out on seven large-scale open-source projects revealed that their approach effectively uncovers how architectural issues evolve into technical debt over time.
Martini \textit{et al.} \citep{martini2018semi} conducted a case study within a large company, establishing a comprehensive framework for the semi-automated identification and estimation of architectural debt. 
In their effort estimation, they considered factors such as the number of files, lines of code, changes in all files, and McCabe's and Halstead's complexity metrics.
Nugroho \textit{et al.} \citep{nugroho2011empirical} proposed an approach to quantify debt in terms of fixing technical quality issues and the extra cost spent on maintenance.
Specifically, they estimated the repayment effort by calculating the rework fraction and rebuild value: the rework fraction represents the percentage of lines of code requiring modification to enhance software quality; the rebuild value estimates the effort (in man-months) necessary to rebuild a system using a particular technology.

Furthermore, two studies have empirically investigated the repayment effort of SATD.
Mensah \textit{et al.} \citep{mensah2018value} analyzed SATD items to identify instances of ``vital few'' (bug-prone) tasks and ``trivial many'' (less bug-prone) tasks.
They used the number of commented lines of code as a measure of effort estimation for SATD.
The results indicated that highly prioritized (vital few) SATD tasks required a rework effort of modifying 10 to 25 commented LOC per source file.
Wehaibi \textit{et al.} \citep{wehaibi2016examining} explored whether SATD changes require more effort to be repaid than non-SATD changes.
They identified SATD and non-SATD changes using 62 SATD keywords \citep{potdar2014exploratory}.
They then compared the difficulty of SATD and non-SATD changes using four measures: the total number of modified lines, files, and directories, and change entropy.
Their results suggested that SATD changes were more challenging than non-SATD changes across all four measures of difficulty.

\textbf{Our study differs from the aforementioned studies} on SATD repayment effort in several aspects.
First, we employ state-of-the-art machine learning techniques to identify SATD items, as opposed to relying on the list of 62 SATD keywords from earlier work \citep{potdar2014exploratory}.
Second, we extract SATD-related changes corresponding to commits identified as SATD items.
Specifically, after employing machine learning techniques to pinpoint commits with SATD items, we further analyze the related code changes to measure the repayment effort involved.
Third, we examine four distinct types of SATD items, rather than treating them as a single, homogeneous group, allowing for more nuanced insights into different types of SATD.
Fourth, we integrate a broader range of metrics to offer a more comprehensive assessment of repayment effort, enhancing our understanding of the SATD repayment.
Fifth, our analysis covers over 1,000 repositories, representing a significant expansion compared to the five repositories investigated in previous studies.
Finally, and most notably, we introduce the first-ever approach for predicting SATD repayment effort based on SATD textual information, as well as identifying keywords associated with varying levels of repayment effort.
This innovative approach offers valuable insights and guidance for developers managing SATD.

\section{Conclusion and Future Work}
\label{sec:conclusion}

In this study, we investigated the effort required to repay SATD using a dataset of 341,740 SATD items from 2,568,728 commits across 1,060 Apache repositories.
Our findings show that SATD and non-SATD items require similar direct resolution effort (mean total lines changed: 131.5 vs.\ 132.7), but SATD repayment creates significantly larger ripple effects, with more medium- and high-significance code changes (effect size = 0.08, p $<$ 0.05).
Different SATD types demand distinct effort levels: documentation debt requires the least effort (72.7 lines), even lower than non-SATD items, while requirement debt demands the most (169.7 lines), highlighting the need to treat each type distinctly when prioritizing repayment.
For automated prediction, BERT achieves the best average RMSE of 13.5 for lines of code prediction, reducing the baseline error (38.1) by 64.6\%, and we further identified interpretable keywords where low-effort items involve terms like \emph{typo} and \emph{unused import}, while high-effort items are associated with \emph{code cleanup} and \emph{interface}.
As future work, we plan to incorporate SATD type information into the prediction models, extend our approach to other documentation sources such as code comments, issue trackers, and pull requests, and investigate SATD repayment effort across different programming languages and project types.

\vspace{2mm}
\section*{Data Availability Statement}
The data and code used in this study are publicly available at \url{https://doi.org/10.5281/zenodo.19252072}.

\newpage
\balance
\bibliographystyle{ACM-Reference-Format}
\bibliography{bibliography}

\end{document}